Modeling the interaction of an arbitrary light field with a diffraction grating by the Monte Carlo method

V. V. Savukov, I. V. Golubenko

A high-accuracy solution of the diffraction problem has become necessary for the treatment of certain special questions of statistical physics. This article reports the creation of a computer program that serves as an instrumental method of calculating the parameters of diffraction phenomena when complex optical systems are being theoretically investigated. The program solves the diffraction problem by a rigorous method based on Maxwell's equations under specified boundary conditions. An arbitrary—for instance, diffuse—configuration of the initial light field is allowed. Reflective gratings with a linear or crossed sinusoidal profile of the surface microrelief are considered as the diffraction optical elements. The characteristics of the self-consistent total light field can be calculated when several diffraction elements are present in the system.

*St. Petersburg, Russia, 2012*

## TABLE OF CONTENTS

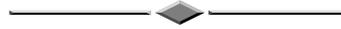

## Introduction

It is necessary in a number of cases of practical significance to solve the diffraction problem presented in extremely general form. For example, the initial light field that interacts with some diffraction optical element (DOE) can have any physically possible parameters. In particular, such primary (before diffraction) radiation can be characterized by a specific spectral composition, as well as by a given angular distribution of the radiance and the degree of polarization in geometrical space. Alternatively, a DOE placed in the initial light field is a reflective diffraction grating made from a designated material with specified macroscopic shape and parameters of the surface microrelief. In a particular case, this surface can be mirror-smooth. If more than one DOE lies in the space under consideration, a self-consistent stationary state of the resulting light field becomes necessary in the calculation.

Such nontrivial problems very rarely can be mathematically described in an exact analytical way. It most often happens that, even if some analytical version exists, it is usually obtained by using fairly arbitrary approximations whose subsequent effect on the adequacy of the created model cannot be objectively estimated.

Methods of simulation modelling are most eminently suitable for solving the indicated problem—in particular, the method of statistical testing known as the Monte Carlo method. The result of solving an extremely complex initial problem in this case is reduced to a statistical generalization of the results of solving a set of certain simpler (more elementary) problems, each of which can already be solved directly [1].

## Formulation of the problem

In considering certain special questions associated with investigating the axiomatic principles of statistical physics [2 - 4], the need arose for high-accuracy implementation of a rigorous method of solving the diffraction problem, based on the solution of Maxwell's equations for specified boundary conditions and which would meet the following requirements:

1. The method must be applicable to the problem of the interaction of an arbitrary light field (in particular, blackbody radiation) with a given diffraction grating.

2. The elementary physical object whose diffraction is to be subsequently analyzed is a monochromatic electromagnetic wave with a planar propagation front. The given wave object can serve as a quasi-classical model of a so-called free photon found in a zone remote from the radiation source. Such a single photon is characterized in each individual case by the following parameters:
   - the wavelength and, as a consequence, the Planck energy;
   - the angle of incidence of the photon on the surface of the diffraction grating (i.e., the angle between the photon's wave vector and the normal to the macroscopic surface);
   - the azimuthal (conical) angle of rotation of the plane of incidence of the photon;
   - the polarization phase (the shift angle between the electric vector and the magnetic induction vector in the wave);
   - the ratio between the sides of the rectangle described around the polarization ellipse.

3. The diffraction grating must be a phase-type reflective grating with one-dimensional (linear) or two-dimensional (crossed) sinusoidal surface microrelief and must have the following numerical parameters:
   - the grating period (step)—separately for each measurement,

- the total depth of the grating microrelief—separately for each measurement,
- the complex permittivity of the actual (i.e., with a finite conductivity) grating material, functionally dependent on the radiation wavelength. The case of ideal conductivity must be described separately.

4. It must be possible to correctly calculate the scattering of each single wave object (photon) on the diffraction grating for the most general case of combinations of the photon and grating parameters. This mainly relates to analyzing the phenomena in the zones of Rayleigh–Wood anomalies, especially for sub-wavelength gratings.

5. The results of calculating the diffraction of a single photon must include the absorption probability of the given photon by the grating and the number of orders (coherent channels) of photon scattering. For each detected order, the efficiency and the angular characteristics of the reflection from the grating are computed, along with the polarization parameters of the radiation in a specific scattering channel.

## The general methodology of the calculations

The required structure of the initial light field (see point 1) is generated by means of probability-based selections of the initial parameters for each separate wave object (see point 2). The results of the solutions (see points 4 and 5) obtained for all the elementary diffraction-scattering events of the photons at the grating (see point 3) are stored by the program. The final parameters of the radiation are found in numerical form as the statistical moments of the distributions, obtained as a consequence of the generalized information of the calculations for all the single elementary scattering events. To make it as clear as possible, the final information is also displayed in the form of graphical images—for example, in the form of angular radiance diagrams of the scattered radiation.

## Calculating a single (elementary) diffraction event

To solve the formulated problem, it was required to consider diffraction at the surface, which possesses periodicity along two axes of a Cartesian coordinate system. Such a surface, which has been called a three-dimensional modulated profile or crossed diffraction gratings, is formed when two two-dimensional modulated profiles or ordinary diffraction gratings are superimposed, each of which is periodic along one of two mutually perpendicular directions.

It is proposed to numerically analyze the properties of such gratings in a wide range of variation of the wavelength-to-period ratio, including the resonance region, in which this ratio approaches unity. The character of the diffraction in the resonance region is substantially affected by the polarization of the incident light. A rigorous method of calculation, based on a solution of the wave equations, or even on a solution of Maxwell's equations, which make it possible to take into account the character of the polarization, can therefore be suitable for numerical analysis.

Rigorous theories on the basis of which computational algorithms for calculating diffraction processes can be constructed are traditionally divided into two main categories: integral and differential. Integral methods assume a numerical solution of integral equations or systems of coupled integral equations, while differential methods produce a solution of infinite systems of coupled differential equations. As shown by practice in the application of the indicated theories, when systems of differential equations are being solved, situations of instability arise in which only the use of integral methods makes it possible to obtain correct results [5]. On the other hand, the possibilities of applying currently existing integral methods are strongly limited because of the complexity of satisfying the boundary conditions for an arbitrary direction of the incident waves. Therefore, it was not a simple problem to choose a method for carrying out the given study.

Much attention is currently being paid to solving the diffraction problem on a surface that possesses double periodicity, and many publications are devoted to it. There are many different ways to solve this problem. Existing rigorous methods include the method of finite differences, the method of boundary variation, and analysis on the basis of coupled waves, as well as various methods of using curvilinear coordinates, of which the most reliable and effective is Chandezon's C-method [6 - 8]. The main feature of the C-method is that it uses a transformation of the coordinate system that converts the grating surface into one of the coordinate surfaces. This conversion is called the translation coordinate system.

By using the tensor form to write Maxwell's equations in the translation coordinate system, which is considered in detail in Ref. [9], it is possible to obtain a linear system of differential equations with constant coefficients. The latter circumstance is very important, since it allows a solution of the diffraction problem to be found by computing the eigenvalues and eigenvectors of the matrix of this system. The given method has a significantly wider region of applicability by comparison with other differential methods and is free from their inherent drawbacks. Transforming the relief surface of the grating to a formally planar form makes it much easier to write the boundary conditions. The method also allows the calculations to be carried out simultaneously for cases of **TM** and **TE** polarization.

To carry out calculations based on the C-method, a numerical algorithm implemented in FORTRAN was developed that makes it possible to perform reliable calculations for various wavelength regions, profile parameters, and various forms of grating material, including a version with infinite conductivity. References [10, 11] served as a prototype for this algorithm, by developing the C-method as applied to gratings with layers of dielectric coatings.

Besides the diffraction efficiency, which is one of the most important characteristics of a grating, the algorithm thus developed can be used to compute the complex amplitudes and phases for **TM** and **TE** polarized diffraction waves. In the final analysis, this makes it possible to determine the elliptical-polarization parameters of each wave—i.e., to determine the shape and slope of the axes of the ellipse.

Besides solving the problem for a three-dimensional modulated metallic surface, the algorithm makes it possible to calculate the scattered light field in the regime of conical diffraction for an ordinary (linear) grating with finite conductivity. Moreover, with insignificant modification, the algorithm can be used for transparent dielectric gratings, as well as gratings with a dielectric coating.

## Structure of the software complex

The software complex created here consists of the following main components:
- The basic part of the program, written in the interpreted language of the **Maple**™ mathematical system, version 15.01 (produced by **Waterloo Maple, Inc.**). This part, which has the form of an "interactive worksheet," contains complete information concerning the formulated problem: the parameters of the initial light field, all the characteristics of the DOEs located in this field, etc. Besides this, the given worksheet has an exhaustive description of what calculational information needs to be accumulated, how this information must be processed, and in what visual form the results of the processing should be presented to the user.

- Program modules in the form of dll libraries, called as needed from the basic part of the complex. These modules, the original texts of which are written in the **Fortran 95** language, contain algorithms for calculating single diffraction events. They can be used for different versions of the microrelief dimension and different conductivities of the reflecting surface of the DOE. The modules are obtained by means of the **Compaq Visual Fortran**™ compiler, version 6.6, and include access to the **IMSL® Fortran Numerical Library**, version 4.01.

- Text files with tables of special multidimensional Sobol' sequences [12, 13], intended for the use of the Monte Carlo method with ultrahigh resolution. These files are accessed from the basic part of the complex when grids of values of the input parameters of the problem are formed. The files of tables have an invariant form for the different versions of the problems to be solved. The given files were therefore generated once by means of the corresponding program in **Pascal**, carried out in the **Borland Delphi**™ medium, version 6.0, from the **Borland Software Corporation**.

All the starting data are thus input and the results are output through an interactive worksheet (the **Maple**™ system) of the basic part of the program complex thus created, functioning on the **Win32®** platform.

## Verifying the program

The condition that energy balance is observed was chosen as the main criterion that makes it possible to monitor the correct operation of the software algorithm. To do this, the energy that reaches the grating along with the incident wave was compared with the energy that is carried away by the propagating diffraction waves or was absorbed in the grating material. When the program operates correctly, energy balance must always be maintained with high accuracy.

However, it is well known that maintaining energy balance cannot serve as the only proof that the diffraction problem is solved correctly. The developed algorithm was therefore thoroughly checked by comparing the results of the calculations with analogous information obtained by integral methods both for ideally conductive gratings and for gratings with finite conductivity [5]. Here, among other things, such special regimes were considered as, for example, the case in which the radiation was completely absorbed by the grating surface [14]. Comparative testing of the data obtained for crossed gratings was also carried out with the earlier results of Ref. [10].

To assure the correctness of the program thus created, the results of its operation were directly checked experimentally in one of the most complex regimes: A test was run of the calculation of the diffraction scattering of monochromatic diffuse radiation (wavelength $\lambda = 470$ nm) at a linear sub-wavelength phase grating made from aluminum (step $d = 372$ nm, total depth of the sinusoidal profile of the microrelief $h = 118$ nm) in the presence of zones of intense development of the Rayleigh–Wood anomalies [15, 16].

According to the described logic of simulation modelling, the given diffuse radiation has a discrete structure (it is a photon gas) and is defined as unpolarized incoherent electromagnetic radiation whose separate photons can have with equal probability any possible angular orientation of their wave vectors **k** in three-dimensional geometrical space. Each photon lies in a zone remote from the source and is therefore characterized by a planar wave front.

Figure 1(a) shows a graphical image of a computer model of the total spatial radiance distribution of the scattered light field. The distribution is constructed in the polar coordinate system in such a way that its center corresponds to zero reflection angle when the surface of the diffraction grating is surveyed. The reflection angle is proportional to the polar radius, and this angle approaches 90° at the periphery of the circular diagram. The azimuthal angle of observation of the grating surface is determined by the polar angle of the diagram.

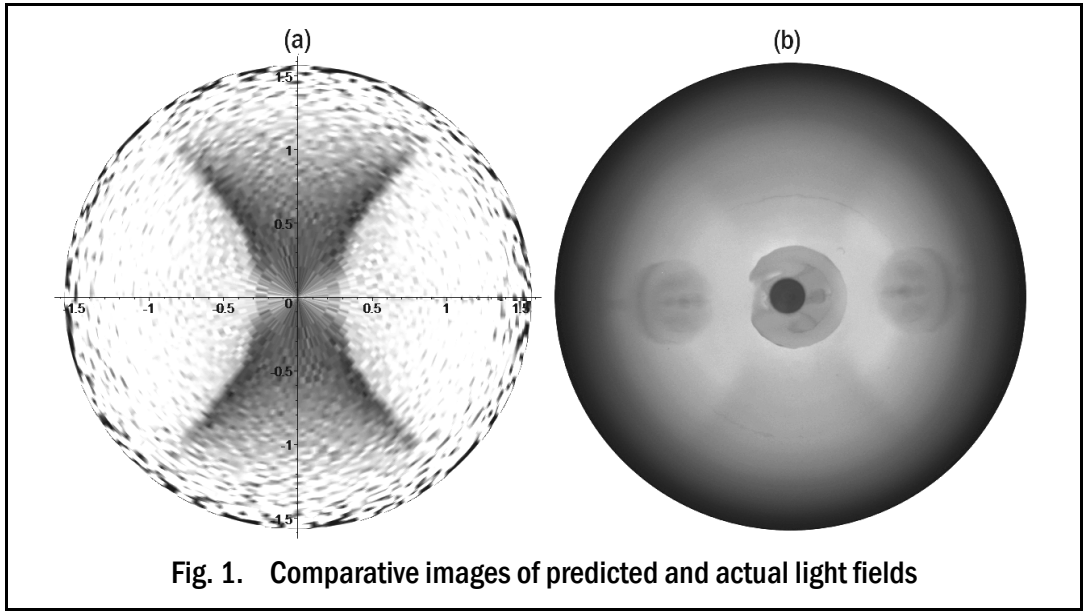

Fig. 1. Comparative images of predicted and actual light fields

Figure 1(b) shows a photograph of the distribution of the scattered light field in an actual physical apparatus, the operation of which was predicted in Fig. 1(a). Here a principle for obtaining a three-dimensional image is used that fixes the parameters of the object for almost all possible combinations of angles of observation at once, known as the fisheye effect. This principle is based on photographing the reflection of the surface of a diffraction grating in a spherical mirror located a short distance from this grating. The picture is taken through a small aperture in the body of the grating itself. Three secondary reflections of the grating surface from the sphere can therefore be seen at the center, the right, and the left on the photograph in Fig. 1(b), and these reflections have no relation to the intrinsic scattering distribution.

The obvious coincidence of the overall form of the angular distribution in Figs. 1(a) and 1(b) is evidence of the successful character of the experimental checking that was carried out.

## Examples of the results of operation

As pointed out earlier, the results of the operation of the software complex described here, which characterize the final parameters of the scattered radiation, can be represented in numerical form and in the form of graphical images. Since the graphical version is more obvious, some angular diagrams of the radiance of the radiation reflected from metallic surfaces with different properties are shown here as an example.

Figures 2 and 3 show graphical images of the distribution of various components of the radiance of the scattered light field for some forms of reflecting surfaces. The initial light field in all cases is monochromatic, diffuse radiation with $\lambda = 555$ nm, and the total number of photons in each statistical experiment is $2^{18} = 262144$. The character of the construction of these distributions is similar to that described earlier for the graph in Fig. 1(a).

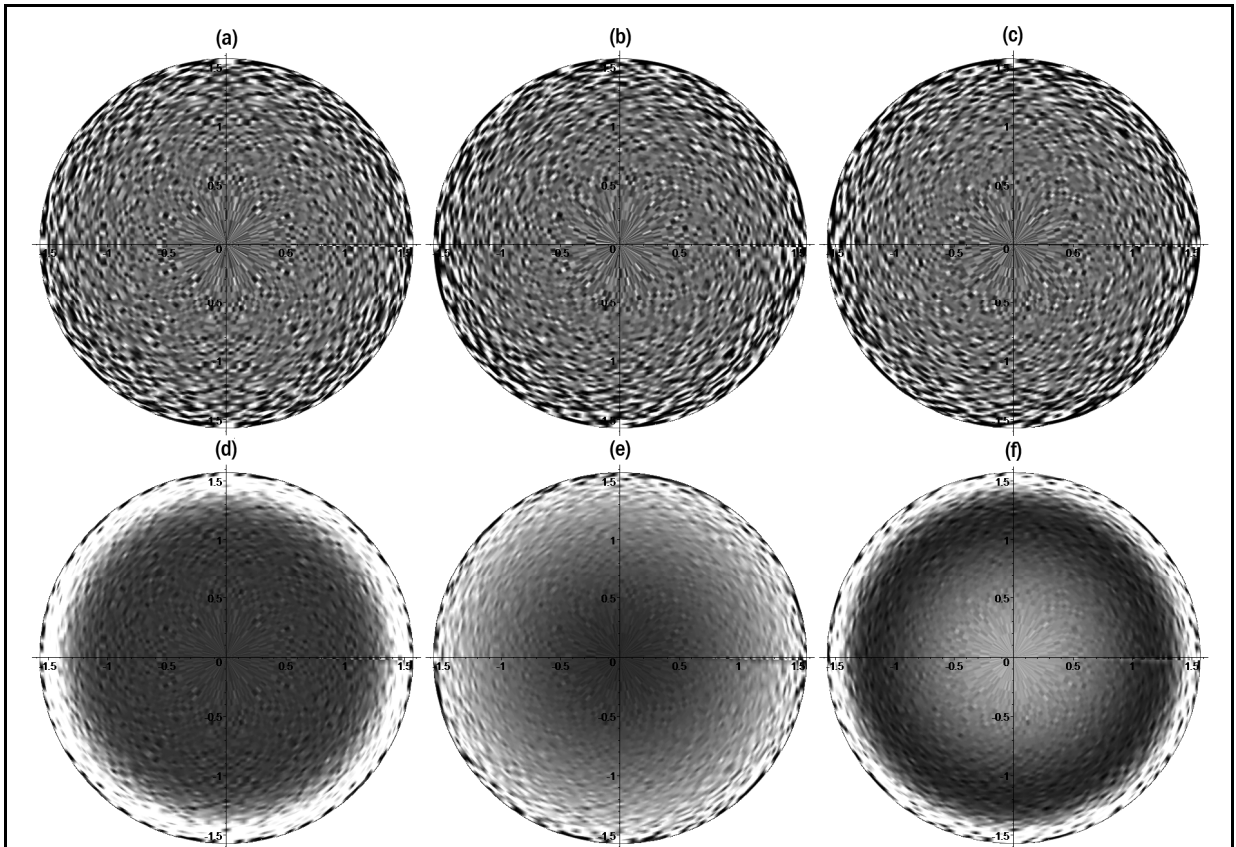

Fig. 2. Surface radiance of ideal (a–c) and copper (d–f) mirrors in diffuse light:
(a) and (d) are graphs of the total radiance of S- and P-polarized reflected components;
(b) and (e) are graphs of the radiance of the S-polarized component of the reflected radiation;
(c) and (f) are graphs of the radiance of the P-polarized component of the reflected radiation.

The distributions in Figs. 2(a)– 2(c) describe the angular radiance distribution of the diffuse light field reflected from an ideal conducting metallic mirror (the S+P, S- and P-polarized components).

Each separate graph is automatically scaled so that it maximally displays all the available contrasts of the scattered light flux density. It is obvious that only unsystematic displays of the scattered field fluctuations that do not form any visually observable macroscopic gradients are present on the images of the distributions in Figs. 2(a)– 2(c). In other words, when diffuse radiation is reflected from an ideally conductive mirror, Lambert's law is satisfied for all the scattered components, and this is quite expected.

The distributions in Figs. 2(d)– 2(f) describe the angular radiance distribution of the diffuse light field reflected from an ideal regular metallic copper mirror. Unlike Figs. 2(a)– 2(c), the images of the distributions in Figs. 2(d)– 2(f) show macroscopic gradients whose scale is automatically increased to the maximum possible value for greater visibility[1]. The fluctuational phenomena in this case become less noticeable on the background of observed macroeffects.

The distributions in Figs. 3(a)– 3(c) describe the angular radiance distribution of the diffuse light field diffraction-scattered at a linear phase grating made from copper (d = 544 nm, h = 337 nm).

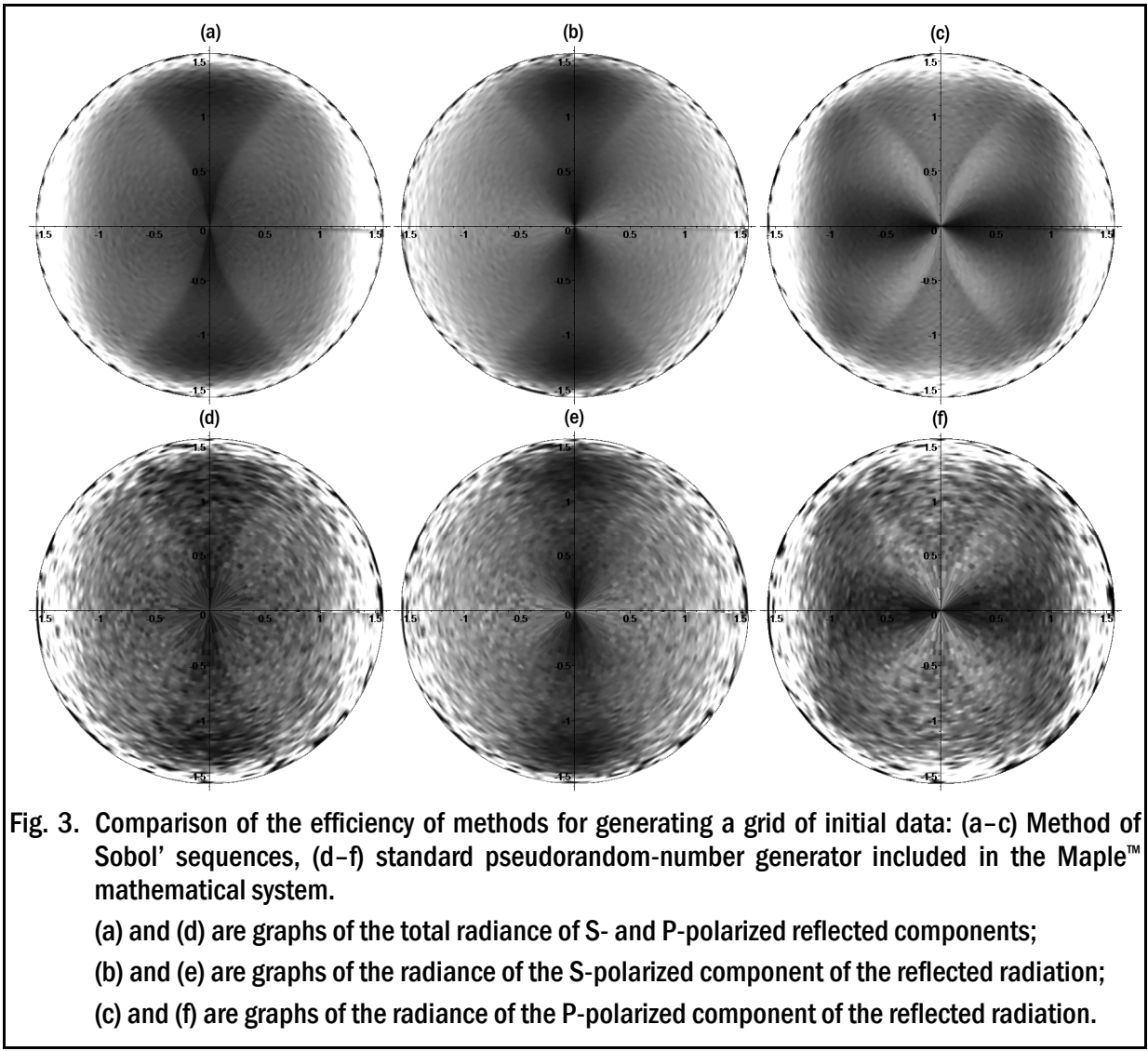

Fig. 3. Comparison of the efficiency of methods for generating a grid of initial data: (a–c) Method of Sobol' sequences, (d–f) standard pseudorandom-number generator included in the Maple™ mathematical system.

(a) and (d) are graphs of the total radiance of S- and P-polarized reflected components;

(b) and (e) are graphs of the radiance of the S-polarized component of the reflected radiation;

(c) and (f) are graphs of the radiance of the P-polarized component of the reflected radiation.

The distributions in Figs. 3(d)– 3(f) describe the same angular radiance distributions as shown in Figs. 3(a)– 3(c). The only difference is that a high-efficiency method of Sobol' sequences was used in Figs. 3(a)– 3(c) to form the values of the network of original data of the statistical tests, while the versions in Figs. 3(d)– 3(f) were obtained using for these purposes the standard pseudorandom-number generator included in the **Maple**™ mathematical system.

---

[1] Their actual scale is computed via the corresponding statistical moments.

Here we do not show the calculated results obtained for a grid with a uniform partitioning step in the region of allowable values of the input parameters, since that type of partitioning, as is well known, is the worst type for multidimensional regions.

## Conclusion

The extensive functional possibilities of the software complex thus created, as well as the fundamentally high reliability of the results obtained with its help, make it possible to effectively use the given tool in theoretical investigations that characterize physical optics. Moreover, it becomes possible in many cases associated with the experimental study of nontrivial optical systems to replace a direct physical experiment with a computer simulation. Such a replacement is most often justified from the viewpoint of economy. Moreover, it gives an advantage in how long it takes to carry out the work.

## Acknowledgment

The authors express deep gratitude to Dr. Daniel Maystre (France) for constructive comments that promoted carrying out this project.

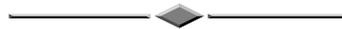